\documentclass[aps,prl,twocolumn,superscriptaddress]{revtex4-1}
\usepackage{amsmath,amssymb,bm}
\usepackage[pdftex]{graphicx}

\begin{document}

\title{Topological Phase Transition in Metallic Single-Wall Carbon Nanotube}

\author{Rin Okuyama}
\email{rokuyama@rk.phys.keio.ac.jp}
\affiliation{Faculty of Science and Technology,
	Keio University, Yokohama 223-8522, Japan}

\author{Wataru Izumida}
\affiliation{Department of Physics, Tohoku University,
	Sendai 980-8578, Japan}

\author{Mikio Eto}
\affiliation{Faculty of Science and Technology,
	Keio University, Yokohama 223-8522, Japan}

\date{December 5, 2016}

\begin{abstract}
The topological phase transition is theoretically studied in
a metallic single-wall carbon nanotube (SWNT) by applying a magnetic
field $B$ parallel to the tube.
The $\mathbb{Z}$ topological invariant, winding number, is
changed discontinuously when a small band gap is closed
at a critical value of $B$, which can be observed as a change
in the number of edge states owing to the bulk-edge correspondence.
This is confirmed by numerical calculations for finite SWNTs of
$\sim 1 \, \mu$m length, using
a one-dimensional lattice model to effectively describe the
mixing between $\sigma$ and $\pi$ orbitals and
spin-orbit interaction,
which are relevant to the formation of the band gap in metallic SWNTs.
\end{abstract}

\maketitle

\noindent
{\it Introduction}---The single-wall carbon nanotube (SWNT) is
a unique one-dimensional (1D) system,
made by rolling up a graphene sheet which possesses two Dirac cones at
$K$ and $K'$ points.
Its helical structure is specified by the chiral vector,
$\bm{C}_{\rm h} = n \bm{a}_1 + m \bm{a}_2 \equiv (n, m)$, where
$\bm{a}_1$ and $\bm{a}_2$ are the
primitive lattice vectors of graphene shown in
Fig.\ 1(a)
	\cite{Saito1998}.
The SWNT is a metal (semiconductor) for $\mod(2n+m,3)=0$ ($\ne 0$) because
some wavevectors discretized in the circumference direction pass
(do not pass)
through the Dirac points when they are expressed in the two-dimensional
(2D) wavevector space of graphene as so-called cutting lines.
Even in a metallic SWNT, a small energy gap is opened between
the conduction and valence bands by the mixing between $\sigma$ and
$\pi$ orbitals owing to a finite curvature of the tube surface
	\cite{Hamada1992, Saito1992, Kane1997}.
The curvature also enlarges the spin-orbit (SO) interaction
	\cite{Ando2000}.

The SWNT can be regarded as a 1D topological insulator because
of the sublattice symmetry for $A$ and $B$ lattice sites
	\cite{Ryu2002}.
It is characterized by a $\mathbb{Z}$ topological invariant,
i.e., winding number,
in both the absence (class BDI) and
presence (AIII) of a magnetic field
	\cite{Wen1989, Schnyder2009}.
The winding number was examined by two of the present
authors for semiconductor SWNTs where neither $\sigma$-$\pi$ mixing
nor SO interaction is relevant to the energy gap
        \cite{Izumida2016}.
It is related to the number of edge states in
finite SWNTs with a suitable boundary condition
by the bulk-edge correspondence.

In this work, we investigate a topological phase transition in
metallic SWNTs by applying a magnetic field $B$ parallel to the tubes.
The small band gap
$E_{\rm g} = 0.1 - 10~{\rm meV}$
at $B=0$ is
closed at the critical magnetic field
$B^* = 1 - 10~{\rm T}$,
where the winding number is changed discontinuously.
This topological phase transition can be observed as different
numbers of
edge states at $B<B^*$ and $B>B^*$, e.g., using scanning tunneling
spectroscopy
	\cite{Kobayashi2005}.
Although a similar phase transition might be possible in semiconductor
SWNTs with $E_{\rm g} \sim 1~{\rm eV}$
	\cite{Ajiki1993},
it would require an unrealistic field of $B^* \sim 10^3~{\rm T}$.

Previously, some groups theoretically proposed that
the magnetic field $B$ changes the number of edge states in metallic SWNTs
of zigzag type, without the consideration of topology
	\cite{Sasaki2005, Marganska2011}.
Here, we examine the winding number and edge states in metallic SWNTs of
general types.
For this purpose, we make a 1D lattice model to effectively describe the
$\sigma$-$\pi$ mixing and SO interaction,
which are relevant to the formation of a small band gap,
as an extension of the effective model proposed in Ref.\
	\cite{Izumida2016}.
Our model is applicable to the metallic SWNTs of length $L_{\rm NT}=$
a few $\mu$m,
which enables the calculation of edge states with a decay
length of $10^2 - 10^3~{\rm nm}$.
This is advantageous compared with
a tight-binding model with all $\sigma$ and $\pi$ orbitals at each carbon atom
	\cite{Izumida2015},
which can be used for $L_{\rm NT} \lesssim 10^2$ nm by moderate computers.

\begin{figure} \begin{center}
\includegraphics{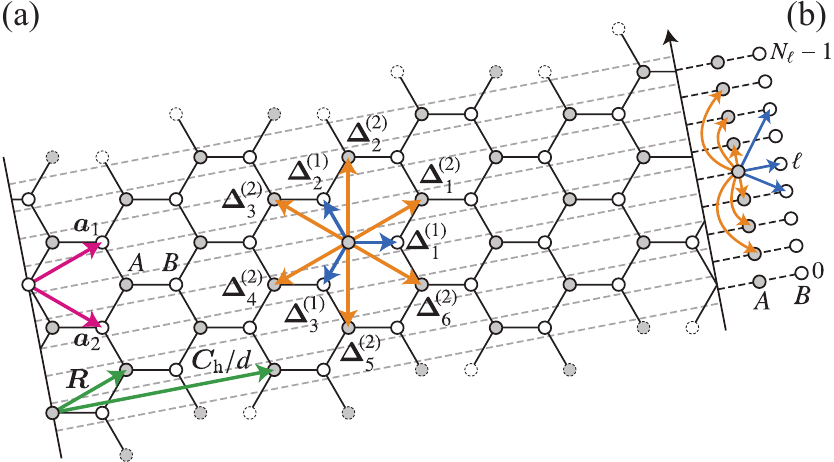}
\caption{\footnotesize
	(a) Primitive lattice vectors of graphene,
	$\bm{a}_1$ and $\bm{a}_2$, and mapping of $(n, m)$-SWNT to the
	graphene sheet.
	The chiral vector,
	$\bm{C}_{\rm h} = n \bm{a}_1 + m \bm{a}_2$,
	indicates the circumference of the tube.
	The three vectors
	$\bm{\Delta}^{\!\!(1)}_j$ ($j=1,2,3$) connect the
	nearest-neighbor atoms, whereas the six vectors
	$\bm{\Delta}^{\!\!(2)}_j$ ($j=1,2,\cdots,6$) connect the
	second nearest-neighbor atoms.
	The rotational symmetry $C_d$ around the tube axis
	(helical symmetry) corresponds to the translational symmetry of
	$\bm{C}_{\rm h}/d$
	($\bm{R} = p \bm{a}_1 + q \bm{a}_2$), where
	$d = \gcd(n, m)$, and $p$ and $q$ are integers
	satisfying $mp - nq = d$
	\cite{note1:helical_vector}.
	This figure shows the case of $(n, m)=(6,3)$ with
	$d=3$, $p=1$, and $q=0$.
	(b) An effective 1D lattice model in which $A$ and $B$ atoms are
	aligned in the axial direction.
	The distant hoppings $\Delta \ell^{(i)}_j$ with phase factor
	$\Delta \nu^{(i)}_j$ of $H_{\mu, s}$ in Eq.\ (\ref{eq:H1d})
	are determined from
	$\bm{\Delta}^{\!\!(i)}_j
	= \Delta \ell^{(i)}_j \bm{R}
	+ \Delta \nu^{(i)}_j (\bm{C}_{\rm h} / d)$
	\cite{note2:helical_vector}.
}
\label{fig:lattice}
\end{center} \end{figure}

\vspace{0.3cm}
\noindent
{\it 1D lattice model}---We construct an effective
1D lattice model for electrons around the
Fermi level $E_{\rm F}$ in a metallic SWNT, starting from the Hamiltonian of
$\bm{k} \cdot \bm{p}$ theory in the continuum limit
    \cite{Izumida2009}.
For $(n, m)$-SWNT, the diameter is given by
$
	d_{\rm t} = |\bm{C}_{\rm h}|/ \pi =
	a \sqrt{n^2 + nm + m^2} / \pi
$
with the lattice constant $a = 0.246~{\rm nm}$ in graphene.
The chiral angle $\theta$ is defined as the angle between
$\bm{C}_{\rm h}$ and $\bm{a}_1$.
For $0 \leq m \leq n$, $0 \leq \theta \leq \pi/6$
with $\theta = 0$ and $\pi/6$ for types of zigzag and armchair,
respectively.
In a magnetic field $B$ in the axial direction, the
Hamiltonian in the vicinity of $K$ and $K'$ points reads
\begin{align}
	\mathcal{H}_{s, \tau} (\bm{k}) &= \hbar v_{\rm F} \Bigl[
		\bigl( k_c - \tau \Delta k_c - s \Delta k_{\rm so}
		- \Delta k_\phi \bigr) \hat{\sigma}_x
		\nonumber \\
	&\quad + \tau \bigl( k_z - \tau \Delta k_z \bigr)
		\hat{\sigma}_y \Bigr]
	+ \Bigl( s \tau \epsilon_{\rm so}
		+ \tfrac12 g_{\rm s} \, \mu_{\rm B} B s \Bigr) \hat{1},
		\label{eq:H_continuum}
\end{align}
where $\hat{\sigma}_x$, $\hat{\sigma}_y$, and $\hat{1}$ are the Pauli
matrices and identity operator, respectively, in the sublattice space
of $\sigma = A$ or $B$.
$s = \pm 1$ is the spin in the axial direction, whereas
$\tau = \pm 1$ is the pseudo-spin to represent the $K$ or
$K'$ valley. $v_{\rm F} = 8.32 \times 10^5~{\rm m/s}$ is the
Fermi velocity and $g_{\rm s} \simeq 2$ is the spin g-factor.
$k_c$ and $k_z$ are the circumference and axial components of
wavenumber measured from $K$ or $K'$ points, respectively:
$k_c$ is discretized in units of
$2\pi/|\bm{C}_{\rm h}|$ while $k_z$ is continuous.

In $\mathcal{H}_{s, \tau} (\bm{k})$,
the hybridization between $\pi$ and $\sigma$ orbitals results
in the shift of Dirac points from $K$ or $K'$ points,
\begin{align}
	\Delta k_c = \beta' \frac{\cos 3\theta}{d_{\rm t}^2}, \quad
	\Delta k_z = \zeta  \frac{\sin 3\theta}{d_{\rm t}^2},
\end{align}
with $\beta' = 0.0436~{\rm nm}$ and $\zeta = -0.185~{\rm nm}$.
$\Delta k_c$ opens a small gap $E_{\rm g}$ between the
conduction and valence bands
around $K$ and $K'$ points,
except $\theta=\pi/6$ (armchair).
The curvature-enhanced SO interaction yields
\begin{align}
	\Delta k_{\rm so} = \alpha'_1 V_{\rm so} \frac{1}{d_{\rm t}}, \quad
	\epsilon_{\rm so} = \alpha_2 V_{\rm so}
		\frac{\cos 3\theta}{d_{\rm t}},
\end{align}
with $\alpha'_1 = 8.8 \times 10^{-5}~{\rm meV^{-1}}$ and
$\alpha_2 = - 0.045~{\rm nm}$. $V_{\rm so} = 6~{\rm meV}$ is
the SO interaction for 2p orbitals in carbon atom.
$\Delta k_{\rm so}$ gives a correction to $E_{\rm g}$.
The Aharonov-Bohm (AB) phase by magnetic field $B$ appears as
\begin{align}
	\Delta k_\phi = - \frac{eB}{4 \hbar} d_{\rm t}.
\end{align}
The band gap is closed at $B^*$ when
$\tau \Delta k_c + s \Delta k_{\rm so} + \Delta k_\phi=0$.
The last term in $\mathcal{H}_{s, \tau} (\bm{k})$ yields the
energy shift from $E_{\rm F}=0$,
which is assumed to be small compared with the band
gap except in the vicinity of $B=B^*$.

We construct a 2D lattice model to reproduce
$\mathcal{H}_{s, \tau} (\bm{k})$ around
the Dirac points
	\cite{supplement}.
The model involves the hoppings not only to the first
nearest-neighbor atoms but also to the second ones.
The former connects $A$ and $B$ atoms, which are depicted by
three vectors $\bm{\Delta}^{\!\!(1)}_j$ ($j=1,2,3$) in
Fig.\ 1(a), whereas the latter connects the same species
indicated by six vectors $\bm{\Delta}^{\!\!(2)}_j$ ($j=1,2,\cdots,6$).

Finally, the effective 1D lattice model is derived from the
2D lattice model, along the lines of Ref.\
	\cite{Izumida2016},
to utilize the helical-angular construction
	\cite{White1993}.
$(n, m)$-SWNT has the $d$-fold symmetry around the tube axis,
where $d = \gcd(n, m)$ is the greatest common
divisor of $n$ and $m$.
It also has the helical symmetry with translation
$a_z = \sqrt3 d a^2 / (2 \pi d_{\rm t})$ along the tube axis with a
rotation around it [see Fig.\ 1(a)].
Thanks to these symmetries, the Hamiltonian is block-diagonalized into
the subspace of orbital angular momentum $\mu = 0, 1, 2, \ldots, d - 1$
and spin $s=\pm 1$,
as $H = \sum_{\mu = 0}^{d - 1} \sum_{s = \pm} H_{\mu, s}$,
\begin{align}
	H_{\mu, s} &= \sum_{\sigma, \ell}
		\frac12 g_{\rm s} \, \mu_{\rm B} B s \,
		c^{\, \mu, s \, \dagger}_{\sigma, \ell}
		c^{\, \mu, s}_{\sigma, \ell}
		\nonumber \\
	& \quad + \sum_{\ell} \sum_{j = 1}^3 \biggl( \gamma^{(1)}_{s, j}
		e^{i 2 \pi \mu \Delta \nu^{(1)}_j / d}
		c^{\, \mu, s \, \dagger}_{A, \ell}
		c^{\, \mu, s}_{B, \ell + \Delta \ell^{(1)}_j}
		+ {\rm h.c.} \biggr)
		\nonumber \\
	& \quad + \sum_{\sigma, \ell} \sum_{j = 1}^3
		\biggl( \gamma^{(2)}_{s, j}
		e^{i 2 \pi \mu \Delta \nu^{(2)}_j / d}
		c^{\, \mu, s \, \dagger}_{\sigma, \ell}
		c^{\, \mu, s}_{\sigma, \ell + \Delta \ell^{(2)}_j}
		+ {\rm h.c.} \biggr).
		\label{eq:H1d}
\end{align}
This is a 1D lattice model in which $A$ and $B$ atoms are aligned in the
axial direction with the lattice constant $a_z$.
$c^{\, \mu, s}_{\sigma, \ell}$ is the field operator of an
electron with angular momentum $\mu$ and
spin $s$ at atom $\sigma$ of site index $\ell$.
The hopping to the first [second] nearest-neighbor atoms in
Fig.\ 1(a) gives rise to the hopping to the sites separated
by $\Delta \ell^{(1)}_j$ [$\Delta \ell^{(2)}_j$] with
phase factor $\Delta \nu^{(1)}_j$ [$\Delta \nu^{(2)}_j$],
as illustrated in Fig.\ 1(b).
The hopping integral $\gamma^{(1)}_{s, j}$
is given by
\begin{align}
	\gamma^{(1)}_{s, j} &= \gamma \biggr[ \exp \left( - i \Delta
		k_\phi a_\text{\tiny CC} \cos \phi^{(1)}_j \right)
		+ \Delta k_c a_\text{\tiny CC} \sin \phi^{(1)}_j
		\nonumber \\
	&\quad - (\Delta k_z + i s \Delta k_{\rm so})
		a_\text{\tiny CC} \cos \phi^{(1)}_j \biggr],
		\label{eq:gamma1}
\end{align}
where
$
	\phi^{(1)}_j = \theta - (5\pi/6) + (2 \pi / 3) j
$
is the angle between $\bm{\Delta}^{(1)}_j$ and $\bm{C}_{\rm h}$,
$a_\text{\tiny CC} = a / \sqrt3$ is the interatomic distance, and
$
	\gamma = - 2 \hbar v_{\rm F} / a_\text{\tiny CC}.
$
$\gamma^{(2)}_{s, j}$
stems from the SO interaction as
\begin{align}
	\gamma^{(2)}_{s, j} = i \frac{(-1)^{j+1}}{3\sqrt{3}}
		s \epsilon_{\rm so}.
		\label{eq:gamma2}
\end{align}

For the bulk states, the Fourier transformation of $H_{\mu, s}$
yields a subband labeled by $\mu$ and $s$ in the first Brillouin
zone, $-\pi/a_z \le k < \pi/a_z$. It is expressed as
\begin{align}
	\epsilon_{\mu, s} (k) &= \epsilon_{\mu, s}^{(0)} (k)
		\pm \bigl| f_{\mu, s} (k) \bigr|,
		\label{eq:subband1} \\
	\epsilon_{\mu, s}^{(0)} (k) &= 
		\frac{2 s \epsilon_{\rm so}}{3 \sqrt3} \sum_{j = 1}^3 (-1)^j
		\sin \left( \frac{2 \pi \mu \Delta \nu_j^{(2)}}{d}
		+ k a_z \Delta \ell_j^{(2)} \right)
		\nonumber \\
		&+ \frac12 g_{\rm s} \, \mu_{\rm B} B s,
		\label{eq:subband2} %\\
\end{align}
\begin{align}
	f_{\mu, s} (k) &= \sum_{j = 1}^3 \gamma^{(1)}_{s, j}
		e^{i 2 \pi \mu \Delta \nu^{(1)}_j / d}
		e^{i k a_z \Delta \ell^{(1)}_j}.
		\label{eq:f}
\end{align}
The system is an insulator when $|\epsilon_{\mu, s}^{(0)} (k)| <
|f_{\mu, s} (k)|$ in the whole Brillouin zone. Then
positive and negative $\epsilon_{\mu, s} (k)$ form the
conduction and valence bands, respectively.
The edge states are obtained by the diagonalization of $H_{\mu, s}$
for a finite system of $N_\ell$ sites ($\ell = 0, 1, \cdots N_\ell - 1$),
or length $(N_\ell - 1) a_z$.

Our effective 1D lattice model is justified as follows.
First, the bulk states in Eq.\ (\ref{eq:subband1}) coincide with those
calculated from the Hamiltonian in Eq.\ \eqref{eq:H_continuum} around
$K$ and $K'$ points. Second, we compare the bulk and edge states
obtained by our model and those by the tight-binding model with all
$\sigma$ and $\pi$ orbitals
	\cite{Izumida2015}
for a system of 50 nm
	\cite{supplement}.
They are in good agreement.

\vspace{0.3cm}
\noindent
{\it Winding number}---The phase of $f_{\mu, s} (k)$ in
Eq.\ \eqref{eq:f} determines the winding number,
\begin{align}
	w_{\mu, s} &= \int_{-\pi/a_z}^{\, \pi/a_z} \frac{dk}{2 \pi}
		\frac{\partial}{\partial k} \arg f_{\mu, s} (k),
	\label{eq:windingnumber}
\end{align}
for subband $(\mu, s)$
	\cite{Izumida2016, Wen1989, Ryu2002, Schnyder2009}.
This is meaningful for an insulator only
	\cite{note3:winding_number}.

The winding number is related to the number of 
edge states, $N_{\rm edge}$, by the bulk-edge correspondence,
\begin{align}
	N_{\rm edge} = 2 \sum_{\mu = 0}^{d - 1} \sum_{s = \pm} |w_{\mu, s}|,
\label{eq:bulk-edge}
\end{align}
when the tube is cut by a broken line in Fig.\ 1(a)
	\cite{Izumida2016}.
The case of the other boundaries is discussed later.
Although the energy levels of edge states are
deviated from $E_{\rm F}=0$,
they are within the band gap as long as the winding number is well-defined.

\begin{figure}[t] \begin{center}
\includegraphics{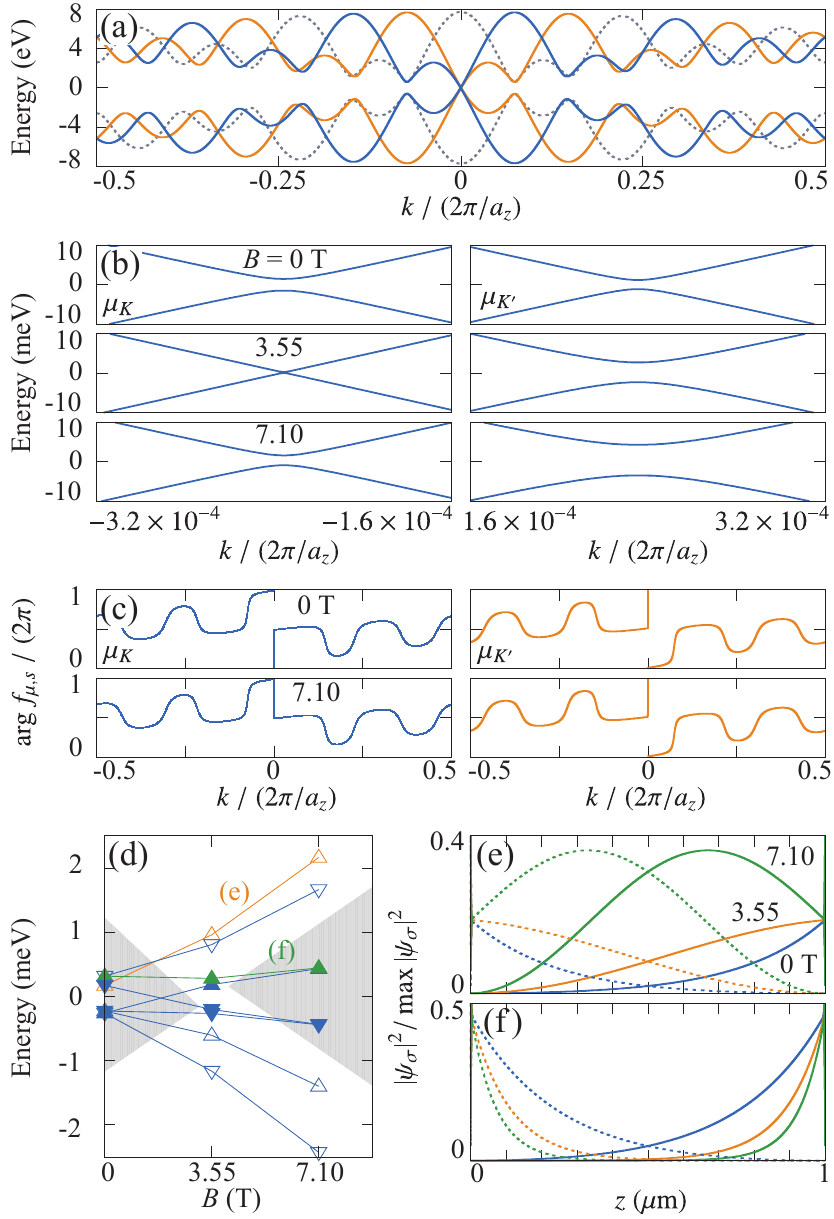}
\caption{\footnotesize
	Numerical results for $(15, 12)$-SWNT.
	(a) Dispersion relation of
	subbands (gray dotted for $\mu=0$,
	orange for $\mu=1=\mu_{K'}$, and blue lines for $\mu=2=\mu_{K}$)
	at magnetic field $B = 0$. They are degenerate for spin
	$s=\pm 1$ in this energy scale.
	(b) Subbands of $(\mu_{K},s=+1)$ and $(\mu_{K'},s=+1)$
	in the vicinity of $k=0$ at $B=0$, 3.55, and 7.10 T.
	(c) $\arg f_{\mu, s} (k)$ in the Brillouin zone for
	subbands of $(\mu_{K},s=+1)$ and $(\mu_{K'},s=+1)$
	at $B=0$ and 7.10 T.
	(d) Energy eigenvalues within
	the band gap at $B = 0$ and
	their $B$ dependence for a finite system of
	$L_{\rm NT} = 1~{\rm \mu m}$.
	The open and filled triangles
	(inverted triangles) correspond to
	$\mu_K$ and $\mu_{K'}$ states, respectively, with $s = +1$ ($-1$).
	Hatched regions indicate the band gap determined from Eq.\
	\eqref{eq:subband1}
	[the band gap is absent between $B^*$s for
	$(\mu_K, s=+1)$ and $(\mu_K, s=-1)$].
	(e), (f) Probability amplitude along the tube axis,
	$|\psi_{\sigma}(z)|^2$ ($\sigma=A, B$), for the upper state
	of (e) $\mu_{K}$ and (f) $\mu_{K'}$ with $s = +1$.
	The solid and dotted lines
	indicate the amplitudes at $A$ and
	$B$ atoms, respectively. $|\psi_{\sigma}(z)|^2$ is
	normalized by its maximum value at a sharp peak around an edge
	(not seen in this length scale).
	\label{fig:15-12}
}
\end{center} \end{figure}

\vspace{0.3cm}
\noindent
{\it Numerical results}---Now we calculate the winding number and
edge states in metallic
SWNTs, using the effective 1D lattice model for finite systems of
$N_\ell$ sites.

The metallic SWNTs with $\mod(2n+m,3)=0$
are categorized to metal-1 or metal-2 according to
the angular momenta of the Dirac points when the tube curvature
and SO interaction are disregarded.
In general, the subband of
$
\mu_{K/K'} = \pm {\rm mod} \bigl[ (2n + m)/3, d \bigr]
$
passes the $K/K'$ point.
$
\mu_K = \mu_{K'} = 0
$
if
$
{\rm mod}(n-m, 3d)=0
$,
$
\mu_K \neq \mu_{K'}
$
otherwise. Metal-1 corresponds to the latter while
metal-2 corresponds to the former.
The wavenumber at the Dirac point is given by 
$k_{K/K'} = \pm (2\pi / 3 a_z)\, {\rm mod}(2p + q, 3)$,
where $p$ and $q$ are integers satisfying $mp - nq = d$.
	\cite{Izumida2016}.

As an example of metal-1, Fig.\ \ref{fig:15-12} presents
the calculated results for $(15, 12)$-SWNT.
For $d = \gcd(15, 12)=3$, we have six subbands with
$\mu=0,1,2$ and $s=\pm 1$.
$\mu_K=2$ and $\mu_{K'}=1$ while $k_K=k_{K'}=0$.
Figure \ref{fig:15-12}(a) shows the subbands in
the whole Brillouin zone when $B=0$ (they seem spin-degenerate in this
energy scale). A small band gap around $k_K=k_{K'}=0$ is given by
$E_{\rm g} = 2.71 \pm s \, 0.32~{\rm meV}$
for $\mu_{K}/\mu_{K'}$ and spin $s=\pm 1$. The band gap of
subband $(\mu_K,s=\pm 1)$ is closed at
$B=B^* = 3.55 + s \, 0.41~{\rm T}$ and reopened at $B>B^*$
while that of $(\mu_{K'},s=\pm 1)$ is always finite, as shown in panel (b).

Figure \ref{fig:15-12}(c) depicts
$\arg f_{\mu, s}$ as a function of wavenumber $k$
for $\mu_{K}$ and $\mu_{K'}$, at $B=0$ and 7.10 T.
At $B = 0$, the behavior of $\arg f_{\mu, s}$ is almost the same
for $\mu_{K}$ and $\mu_{K'}$.
They rapidly but continuously change around $k = 0$ and
increase by $2\pi$ as $k$ runs through the whole Brillouin zone.
Hence Eq.\ (\ref{eq:windingnumber}) yields $w_{\mu, s} = 1$.
At $7.10~{\rm T}$, on the other hand,
$\arg f_{\mu, s}$ behaves differently for
$\mu_{K}$ and $\mu_{K'}$: $w_{\mu, s} = 0$ for $\mu_{K}$
and 1 for $\mu_{K'}$.
This clearly indicates a topological phase transition
at $B=B^*$ for subbands $(\mu_{K},s=\pm 1)$.
For subbands ($\mu=0$, $s=\pm 1$), we always find
$w_{\mu, s} = 0$ (not shown).

Accompanied by this phase transition, the number of
edge states changes from 8 to 4 at $B=B^*$ according to Eq.\
(\ref{eq:bulk-edge}). Figure \ref{fig:15-12}(d) shows eight
energy-eigenvalues within the band gap at $B=0$ and their $B$
dependence. Four of them ($\mu_K$, $s=\pm 1$) go out of the band gap
and merge into the bulk states in conduction or valence bands at $B > B^*$,
whereas the others
($\mu_{K'}$, $s=\pm 1$) stay within the band gap.

Four states of $\mu_K$ and $s=\pm 1$ are changed from edge
to bulk states. Figure \ref{fig:15-12}(e) depicts the
probability amplitude of the upper state of $\mu_K$ and $s=1$ along
the tube axis.
It is localized at the edges (large amplitude at $A$ sites around an
edge and at $B$ sites around the other edge) at $B=0$, whereas it
is delocalized through the tube at $B=7.10$ T.
The states of $\mu_{K'}$ remain localized at the edges,
as shown in Fig.\ \ref{fig:15-12}(f).

\begin{figure}[t] \begin{center}
\includegraphics{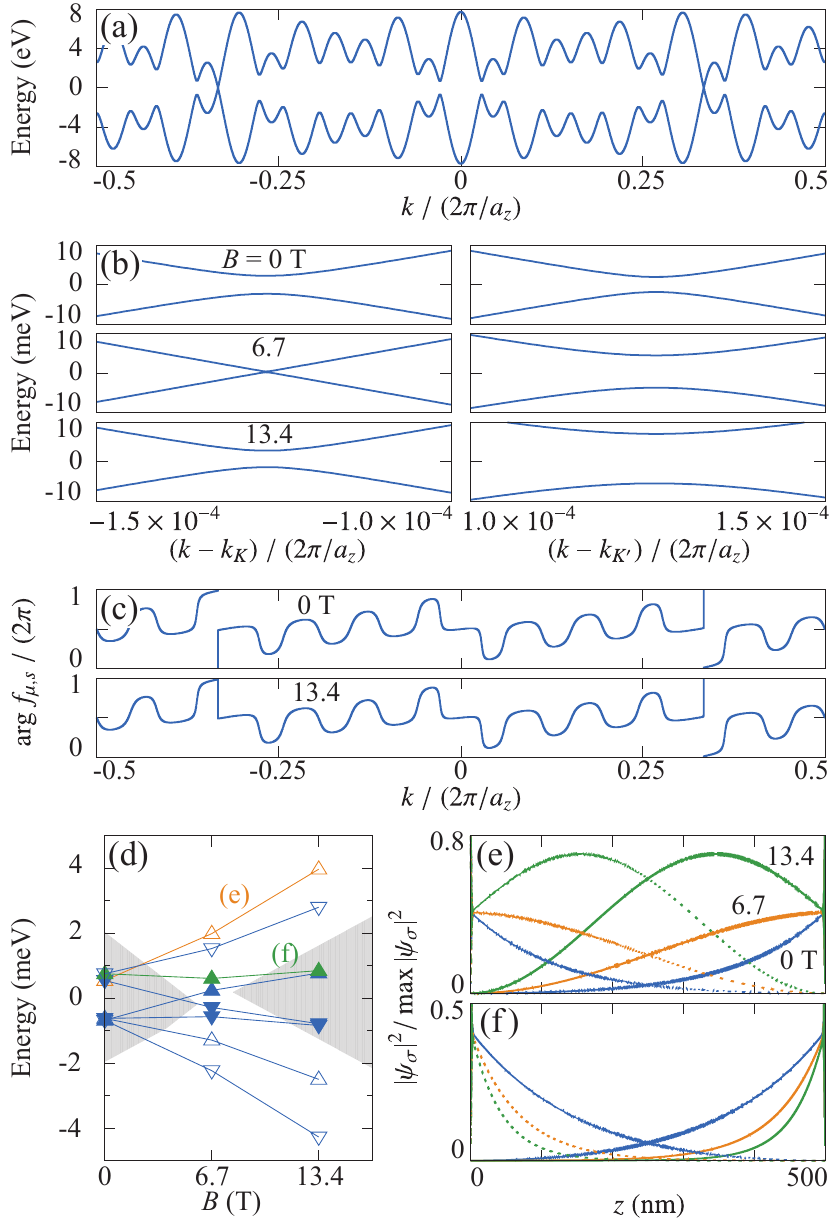}
\caption{\footnotesize
	Numerical results for $(13, 10)$-SWNT.
	(a) Dispersion relation of
	subbands ($\mu=0$, $s = \pm 1$) at
	magnetic field $B = 0$. They seem degenerate in this energy scale.
	A small band gap is opened around $k_K=-2\pi/3$ and $k_{K'}=2\pi/3$.
	(b) Subband ($\mu=0$, $s =1$) in the vicinity of $k = k_K$ and
	$k_{K'}$ at $B = 0$, 6.7, and 13.4 T.
	(c) $\arg f_{\mu, s} (k)$ in the Brillouin zone for subband
	($\mu=0$, $s =1$) at $B = 0$ and 13.4 T.
	(d) Energy eigenvalues within
	the band gap at $B = 0$ and
	their $B$ dependence for a finite system of
	$L_{\rm NT} = 500~{\rm nm}$.
	The open and filled symbols
	correspond to $K$ and $K'$ valley states, respectively,
	for which more than $98\%$ probability amplitude is distributed
	within $|k - k_{K/K'}| < 0.01~(2 \pi / a_z)$ in the wavenumber space.
	Triangles (inverted triangles) represent the states with
	$s = +1$ $(-1)$.
	Hatched regions indicate the band gap determined from Eq.\
	\eqref{eq:subband1}
	(the band gap is absent between $B^*$s for $s=+1$ and $-1$).
	(e), (f) Probability amplitude along the tube axis,
	$|\psi_{\sigma}(z)|^2$ ($\sigma=A, B$), for the upper state
	in (e) $K$ and (f) $K'$ valley with $s = +1$.
	The solid and dotted lines
	indicate the amplitudes at $A$ and $B$ atoms, respectively.
	$|\psi_{\sigma}(z)|^2$ is normalized by its maximum value at a sharp
	peak around an edge (not seen in this length scale).
	\label{fig:13-10}
}
\end{center} \end{figure}

For metal-2, numerical results for $(13, 10)$-SWNT
are given in Fig.\ \ref{fig:13-10}.
For $d = \gcd(13, 10)=1$, we have two subbands of
$\mu=0$ and $s=\pm 1$.
$\mu_K=\mu_{K'}=0$ while $k_K=-k_{K'}=-2\pi/3$.
The band gap is given by $E_{\rm g} = 4.36 \pm s \, 0.37~{\rm meV}$ around 
$k_K$ and $k_{K'}$ at $B=0$. The band gap around $k_K$
is closed at $B=B^* = 6.71 + s \, 0.57~{\rm T}$
while that around $k_{K'}$ is always finite
[panel (b)].

The winding number changes from two to one at $B=B^*$, as seen
in the behavior of $\arg f_{\mu, s}$ in panel (c).
This topological phase transition changes the number of edge states
from 8 to 4: Four of eight states within the band gap at $B=0$
go out of the band gap around $B=B^*$ and the rest remain within the band
gap, as shown in panel (d).
The former change from
edge states to bulk states around $B=B^*$,
whereas the latter are always edge states [panels (e) and (f)].

\noindent
{\it Summary and comments}---We have studied the topological phase
transition in metallic SWNTs by magnetic field $B$ parallel to the tubes,
using an effective 1D lattice model to describe the
mixing between $\sigma$ and $\pi$ orbitals and spin-orbit interaction.
We have demonstrated that a change in winding number is accompanied
by a change in the number of edge states in
$(15, 12)$- and $(13, 10)$-SWNTs, as examples of metal-1 and
metal-2, respectively.

Besides these SWNTs,
we have examined several metallic SWNTs of other chiralities.
The topological phase transition takes place in all the SWNTs of metal-1
that we have studied.
The phase transition is also commonly seen in the case of metal-2 except
for SWNTs of armchair type. In armchair SWNTs, the band gap is opened
by the SO interaction only since $\Delta k_c=0$ in Eq.\ (2).
The closing of the band gap is not accompanied by
the change in winding number
in such SWNTs, which will be explained elsewhere.

A comment should be made on the boundary condition,
which is important for the edge states in 1D topological insulators.
Our numerical calculations have been performed for finite systems
in which a SWNT is cut by a broken line in Fig.\ 1(a)
(angular momentum $\mu$ is
a good quantum number in this case).
This is a ``minimal boundary edge,''
where every atom at the ends has just one dangling bond
	\cite{Akhmerov2008}.
The relation in Eq.\ (\ref{eq:bulk-edge}) holds only for such edges.
Some other boundary conditions result in different numbers of edge states,
as discussed in Ref.\ \cite{Izumida2016}. Even in these cases,
we find the change in the number of edge states at the topological phase
transition. We speculate that this statement is true in general with
any boundaries and also in the presence of impurities inside the bulk.
However, this problem would require further studies. Note that our 1D
lattice model is useful for numerical studies on this problem.

\acknowledgements

The authors acknowledge fruitful discussion with A.\ Yamakage,
K.\ Sasaki, and R.\ Saito.
This work was partially supported by JSPS KAKENHI Grant Numbers
JP16H01046, JP15H05870, JP15K05118, JP15KK0147, and JP26220711.

\begin{widetext}

\newpage
\appendix

\begin{center}
{\bf \fontsize{11pt}{0pt} \selectfont
	Supplemental Material for ``Topological Phase Transition in
	Single-Wall Carbon Nanotube''
}
\end{center}

\begin{center}
	Rin Okuyama,$^1$ Wataru Izumida,$^2$ and Mikio Eto$^1$
\end{center}

\begin{center}
{\it \small
	$^1$ Faculty of Science and Technology, Keio University,
	Yokohama 223-8522, Japan
}\\
{\it \small
	$^2$ Department of Physics, Tohoku University,
	Sendai 980-8578, Japan
}
\end{center}

\section{Derivation of the effective one-dimensional lattice model
	\label{app:derivation}}

First, we derive the effective 2-dimensional (2D)
lattice model for a metallic single-wall carbon nanotube (SWNT),
starting from the Hamiltonian of $\bm{k} \cdot \bm{p}$
theory in Eq.\ \eqref{eq:H_continuum} in the main material.
We do not consider a magnetic field in the axial direction $B$
at the beginning.
We split Hamiltonian \eqref{eq:H_continuum} into two parts,
one originates from the zone-folded graphene
and the other includes the effects of finite curvature and
spin-orbit (SO) interaction,
$
	\mathcal{H}_{s, \tau} (\bm{k}) = \mathcal{H}^{(0)}_{s, \tau} (\bm{k})
		+ \mathcal{H}'_{s, \tau}
$.
The matrix elements are given by
\begin{align}
	\bigl[ \mathcal{H}^{(0)}_{s, \tau} (\bm{k}) \bigr]_{AB} &=
		\chi_\tau \hbar v_{\rm F} (k_c - i \tau k_z),
		\label{eq:mathcal_H_AB} \\
	\bigl[ \mathcal{H}^{(0)}_{s, \tau} (\bm{k}) \bigr]_{AA} &=
	\bigl[ \mathcal{H}^{(0)}_{s, \tau} (\bm{k}) \bigr]_{BB} = 0, \\
	\bigl[ \mathcal{H}'_{s, \tau} \bigr]_{AB} &=
		\chi_\tau \hbar v_{\rm F} (- \tau \Delta k_c
			- s \Delta k_{\rm so}
			+ i \Delta k_z),
		\label{eq:mathcal_H_AB_prime} \\
	\bigl[ \mathcal{H}'_{s, \tau} \bigr]_{AA} &=
	\bigl[ \mathcal{H}'_{s, \tau} \bigr]_{BB} =
		s \tau \epsilon_{\rm so},
		\label{eq:mathcal_H_AA_prime}
\end{align}
with a phase factor $\chi_\tau$ by the gauge degrees of freedom.
$\bigl[ \mathcal{H}^{(0)}_{s, \tau} (\bm{k}) \bigr]_{AB}$ is reduced
from the tight-binding model of graphene with the nearest-neighbor hopping,
\begin{align}
	H^{(0)} = \sum_{s, \bm{r}_A} \sum_{j = 1}^3 \biggl( \gamma
		c^{\, s \, \dagger}_{\bm{r}_A}
		c^{\, s}_{\bm{r}_A + \bm{\Delta}^{(1)}_j}
		+ {\rm h.c.} \biggr),
		\label{eq:H0}
\end{align}
where $\bm{r}_\sigma$ is a position of $\sigma$ atom
and $c^{\, s}_{\bm{r}}$ is the field operator for
the atomic orbital at site $\bm{r}$ with spin $s$.
Indeed its Fourier transform yields
$
	\bigl[ \mathcal{H}^{(0)}_{s, \tau} (\bm{k}) \bigr]_{AB} = \gamma
		\sum_{j = 1}^3 e^{i (\tau \bm{K}
			+ \bm{k}) \cdot \bm{\Delta}^{(1)}_j}
	\simeq \tau e^{i \tau \theta} \hbar v_{\rm F}
		(k_c - i \tau k_z)
$
in the vicinity of $K$ and $K'$ points.
Here, we choose wavenumbers at $K/K'$ points as
$
	\pm \bm{K} = \pm (2 \bm{b}_1 + \bm{b}_2) / 3
$
with $\bm{b}_1$ and $\bm{b}_2$ being the reciprocal lattice vectors
conjugate to $\bm{a}_1$ and $\bm{a}_2$, respectively.
This choice corresponds to
$
	\chi_\tau = \tau e^{i \tau \theta}
$
in Eqs.\ \eqref{eq:mathcal_H_AB} and \eqref{eq:mathcal_H_AB_prime}.
Since the curvature effects are relevant only around $K$ and $K'$ points,
we extrapolate $\mathcal{H}'_{s, \tau}$ into the whole wavenumber space as
\begin{align}
	H' = \sum_{s,\tau,\sigma,\sigma',\bm{k}}
		e^{- \xi^2 (\bm{k} - \tau \bm{K})^2}
		\bigl[ \mathcal{H}'_{s, \tau} \bigr]_{\sigma, \sigma'}
		c^{\, s \, \dagger}_{\sigma, \bm{k}}
		c^{\, s}_{\sigma', \bm{k}}.
		\label{eq:H_prime_k}
\end{align}
Here, we have introduced the cut-off parameter $\xi$;
Eq.\ \eqref{eq:H_prime_k} reproduces
Eqs.\ \eqref{eq:mathcal_H_AB_prime} and
\eqref{eq:mathcal_H_AA_prime} for
$|\bm{k} - \tau \bm{K}| \lesssim \xi^{-1}$.
$c^{\, s}_{\sigma, \bm{k}}$ is the field operator for
a Bloch orbital at $\sigma$ atom with spin $s = \pm 1$.
The Fourier transformation into the real space yields
\begin{align}
	H' &= \sum_{s,\sigma,\sigma',\bm{r}_\sigma,\bm{r}_{\sigma'}}
		\gamma'_{s,\sigma,\sigma'} (\bm{r}_{\sigma'} - \bm{r}_\sigma)
		c^{\, s \, \dagger}_{\bm{r}_\sigma}
		c^{\, s}_{\bm{r}_{\sigma'}},
		\label{eq:H_prime} \\
	\gamma'_{s,\sigma,\sigma'} (\bm{\Delta}) &=
		\frac{a^2 e^{- \bm{\Delta}^2 / (4 \xi^2)}}
			{2 \sqrt3 \pi \xi^2}
		\sum_\tau e^{- i \tau \bm{K} \cdot \bm{\Delta}}
		\bigl[ \mathcal{H}'_{s, \tau} \bigr]_{\sigma, \sigma'}.
\end{align}
The hopping integral $\gamma'_{s, \sigma, \sigma'} (\bm{\Delta})$
rapidly decreases with $|\bm{\Delta}|$.
For the on-site term, we have
$
	\gamma'_{s,\sigma,\sigma}(\bm{0}) = 0
$.
The 1st and 2nd nearest-neighbor terms yield
\begin{align}
	\gamma'_{s, A, B} \left( \bm{\Delta}^{(1)}_j \right)
	&= A_1 \cdot \hbar v_{\rm F}
	\left[ - \Delta k_c \sin \phi^{(1)}_j
	+ (\Delta k_z + i s \Delta k_{\rm so}) \cos \phi^{(1)}_j
	\right],
		\label{eq:gamma1_prime} \\
	\gamma'_{s, \sigma, \sigma} \left( \bm{\Delta}^{(2)}_j \right)
	&= A_2 \cdot i (-1)^{j + 1} s \epsilon_{\rm so},
\end{align}
respectively, with
$
	A_1 = a^2 e^{- a^2 / (12 \xi^2)}
		/ \bigl( \sqrt3 \pi \xi^2 \bigr)
$
and
$
	A_2 = a^2 e^{- a^2 / (4 \xi^2)} / ( 2 \pi \xi^2 )
$.
We truncate the summation in Eq.\ \eqref{eq:H_prime} up to the
2nd nearest-neighbor atoms and treat $A_1$ and $A_2$ as
independent fitting parameters.
By using Eq.\ \eqref{eq:gamma1_prime}, we obtain
$
	\bigl[ \mathcal{H}'_{s, \tau} \bigr]_{AB} \simeq
	\sum_{j = 1}^3 \gamma'_{s, A, B} \left( \bm{\Delta}^{(1)}_j \right)
		e^{i \tau \bm{K} \cdot \bm{\Delta}^{(1)}_j}
		= \tfrac32 A_1 \cdot \chi_\tau \hbar v_{\rm F}
		(- \tau \Delta k_c - s \Delta k_{\rm so}
		+ i \Delta k_z).
$
We then have $A_1 = 2/3$.
In a similar manner, we obtain $A_2 = 1 / \bigl( 3\sqrt3 \bigr)$.
$\xi \sim 0.4~a$ approximately reproduces $A_1$ and $A_2$
simultaneously.

Next, we consider the magnetic field $B$.
The spin-Zeeman term,
$
	\sum_{s, \sigma, \bm{r}_\sigma}
		\frac12 g_{\rm s} \, \mu_{\rm B} B s \,
		c^{\, s \, \dagger}_{\bm{r}_\sigma}
		c^{\, s}_{\bm{r}_\sigma}
$,
is added.
For the orbital magnetism, we take into account the
Aharonov-Bohm (AB) effect only, neglecting a small
deformation of atomic orbitals.
When an electron circulates around the tube axis,
it acquires the AB phase
$
	2 \pi B S (-e) / h = - \pi e B d_{\rm t}^2 / (4 \hbar)
$.
The hopping from a $B$ atom to a nearest $A$ atom corresponds to
the rotation of
$
	- \bm{\Delta}^{(1)}_j \cdot \bm{C}_{\rm h} / |\bm{C}_{\rm h}|^2
	= - a_\text{\tiny CC} \cos \phi^{(1)}_j / \bigl( \pi d_{\rm t} \bigr)
$
around the axis. Therefore, we substitute
$
	\gamma \rightarrow \gamma \exp \left( - i \Delta k_\phi
		a_\text{\tiny CC} \cos \phi^{(1)}_j \right)
$
in Eq.\ \eqref{eq:H0} with $\Delta k_\phi = - e B d_{\rm t} / (4 \hbar)$.
Then, we have
$
	\bigl[ \mathcal{H}^{(0)}_{s, \tau} (\bm{k}) \bigr]_{AB}
	\simeq \chi_\tau \hbar v_{\rm F} (k_c - \Delta k_\phi
		- i \tau k_z)
$.

Finally, we obtain the Hamiltonian,
\begin{align}
	H &= \sum_{s, \sigma, \bm{r}_\sigma}
		\frac12 g_{\rm s} \, \mu_{\rm B} B s \,
		c^{\, s \, \dagger}_{\bm{r}_\sigma}
		c^{\, s}_{\bm{r}_\sigma}
	+ \sum_{s, \bm{r}_A} \sum_{j = 1}^3
		\biggl( \gamma^{(1)}_{s, j} \,
		c^{\, s \, \dagger}_{\bm{r}_A}
		c^{\, s}_{\bm{r}_A + \bm{\Delta}^{(1)}_j} + {\rm h.c.} \biggr)
	+ \sum_{s, \sigma, \bm{r}_\sigma} \sum_{j = 1}^3
		\biggl( \gamma^{(2)}_{s, j} \,
		c^{\, s \, \dagger}_{\bm{r}_\sigma}
		c^{\, s}_{\bm{r}_\sigma + \bm{\Delta}^{(2)}_j}
		+ {\rm h.c.} \biggr).
		\label{eq:H2d}
\end{align}
Note that for $B = 0$, SWNTs have the $C_2'$ symmetry.
In addition, zigzag and armchair SWNTs have two mirror planes
which include the tube axis and are perpendicular to it.
	$\!^{1)}$
The Hamiltonian in Eq.\ \eqref{eq:H1d} satisfies these symmetries.

Now we derive the effective 1D model,
based on the helical-angular construction.
	$\!^{2)}$
Since $(\bm{R}, \bm{C}_{\rm h} / d)$ is a set of
primitive lattice vectors of graphene,
a position of $\sigma$ atom can be denoted by
$
	\bm{r}_\sigma = \bm{r}_\sigma (\ell, \nu) = \ell \bm{R} +
		\nu (\bm{C}_{\rm h} / d) + \delta_{\sigma,B} \bm{\Delta}^{(1)}_1
$
on a 2D plane of graphene,
with site index $\ell$ and $\nu = 0, 1, \ldots, d - 1$.
Here, $A$ and $B$ atoms connected by $\bm{\Delta}^{(1)}_1$ have
the same integer coordinates $(\ell, \nu)$.
The cutting lines are given by
\begin{align}
	\bm{k} = \mu \frac{\bm{Q}_1}{d} + k \frac{\bm{Q}_2}{2 \pi / a_z}, \quad
	(\mu = 0, 1, \ldots,~ d - 1;~ - \pi / a_z \leq k < \pi / a_z)
	\label{eq:cutting_line}
\end{align}
where
$
	\bm{Q}_1 = - q \bm{b}_1 + p \bm{b}_2
$
and
$
	\bm{Q}_2 = (m / d) \bm{b}_1 + (n / d) \bm{b}_2
$
are the reciprocal lattice vectors conjugate to
$\bm{C}_{\rm h}/d$ and $\bm{R}$, respectively.
$\mu$ and $k$ are the orbital angular momentum and
wavenumber in the axial direction.
By performing the partial Fourier transformation
along $\bm{Q}_1$ direction to a linear combination of atomic orbitals,
we obtain eigenstates of angular momentum,
\begin{align}
	c^{s, \, \mu}_{\sigma, \ell} = \frac{1}{\sqrt d} \sum_{\nu = 0}^{d - 1}
		e^{- i 2 \pi \mu \nu / d}
		c^{\, s}_{\bm{r}_\sigma (\nu, \ell)}.
		\label{eq:c_mu}
\end{align}
Substituting Eq.\ \eqref{eq:c_mu} into Eq.\ \eqref{eq:H2d},
we obtain the 1D lattice model in Eq.\ \eqref{eq:H1d} in the main material.

\section{Comparison between the effective one-dimensional model and
	extended tight-binding model
	\label{app:comparison}}

We compare the calculated results for a metallic SWNT by
the effective 1D lattice model and extended tight-binding model (ETB)
with all $\sigma$ and $\pi$ orbitals at each carbon atom.
	$\!^{3)}$
In ETB, hopping and overlap integrals are taken into account between
atoms within the distance of $10\,a_{\rm B}$, where $a_{\rm B}$ is
the Bohr radius.
Their values are evaluated by the {\it ab initio} calculations.
	$\!^{4)}$
The optimization of atomic positions is also performed.
Each dangling bond at the ends is terminated by a hydrogen atom.
Here, we consider (7, 4)-SWNT, which is a metal-2 nanotube with
diameter $d_{\rm t} = 0.76~{\rm nm}$.
Figure \ref{fig:7-4}(a) shows the energy eigenvalues, $\epsilon_i$,
obtained by the both methods for a tube with length $L_{\rm NT} = 50~{\rm nm}$
and minimal boundary edges at both ends.
The magnetic field is $B = 10~{\rm T}$.
Here, $i$ is an index of energy eigenstates in ascending order,
with $i = 0$ and $1$ being the highest occupied molecular orbital (HOMO)
and lowest unoccupied molecular orbital (LUMO), respectively.
The origin of energy is chosen so as to $(\epsilon_0 + \epsilon_1)/2 = 0$.
Triangles and inverted triangles indicate spin-up ($s = +1)$ and
down ($s = -1$) in the axial direction, respectively.
Precisely speaking, an energy eigenstate is not an eigenstate of
electron spin in ETB.
However, the probability amplitude of $s = +1$ $(-1)$ is more than
$99.9999~\%$ in spin-up (-down) states shown in Fig.\ \ref{fig:7-4}(a).
The results by the two methods are in agreement semiquantitatively.
Especially, eight edge states in the energy gap are observed by
the both methods.
The effective 1D model overestimates the energy for the valence band electrons.
This is because the overlap integrals are not taking account,
which enhances the width of valence band.
The spin configuration in the edge states are partially inconsistent,
however these states are almost degenerate.
Figures \ref{fig:7-4}(b) and (c) show the probability amplitude for
$i = 0$ and $5$ states, respectively.
The shapes of edge state ($i = 0$) in panel (b)
and traveling mode ($i = 5$) in panel (c)
show good agreement between these calculations,
respectively.

{\footnotesize
\renewcommand{\labelenumi}{\arabic{enumi})}
\begin{enumerate}
\item
E.\ B.\ Barros, A.\ Jorio, G.\ G.\ Samsonidze, R.\ B.\ Capaz,
A.\ G.\ S.\ Filho, J.\ M.\ Filho, G.\ Dresselhaus, and M.\ S.\ Dresselhaus,
Phys.\ Rep.\ {\bf 431}, 261 (2006).
\item
C.\ T.\ White, D.\ H.\ Robertson, and J.\ W.\ Mintmire,
Phys.\ Rev.\ B {\bf 47}, 5485 (1993).
\item
W.\ Izumida, K.\ Sato, and R.\ Saito,
J.\ Phys.\ Soc.\ Jpn.\ {\bf 78}, 074707 (2009).
\item
D.\ Porezag, T. Frauenheim, T. K\"ohler, G.\ Seifert, and R.\ Kaschner,
Phys.\ Rev.\ B {\bf 51}, 12947 (1995).
\end{enumerate}
}

\begin{figure*}[b] \begin{center}
\includegraphics{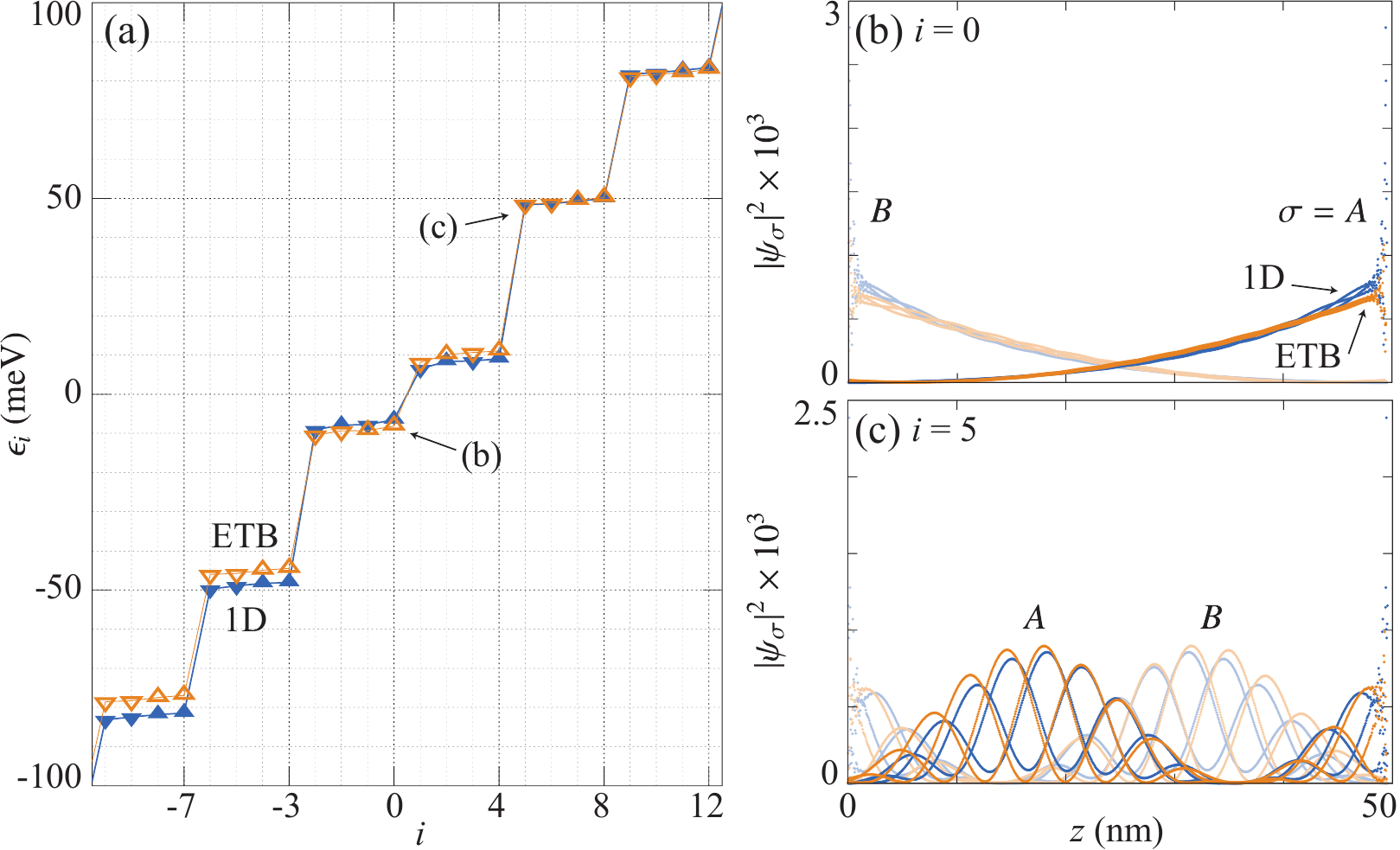}
\caption{\footnotesize
	(a) Energy eigenvalues for (7, 4)-SWNT with
	length $L_{\rm NT} = 50~{\rm nm}$
	at $B = 10~{\rm T}$, calculated by the effective 1D model and ETB.
	Triangles (inverted triangles) indicate $s = +1$ $(-1)$ states.
	(b), (c) Probability amplitude of (b) $i = 0$ and (c) $5$ states,
	respectively.
	The colored (translucent) symbols indicate
	the amplitude on $A$ ($B$) atoms.
	\label{fig:7-4}
}
\end{center} \end{figure*}

\end{widetext}


\begin{thebibliography}{10}
\bibitem{supplement}
Derivation of the effective 1D lattice model and
comparison with the extended tight-binding model are provided online.
\bibitem{note1:helical_vector}
The helical symmetry means the invariance of $(n,m)$-SWNT by
the translation $a_z$ along the tube axis with rotation $\theta_R$
around it. $\theta_R=2 \pi (t_1 q - t_2 p) / N$, where
$t_1 = (2 m + n) / d_{\rm R}$, $t_2 = -(2n + m) / d_{\rm R}$,
and $N = 2(n^2 + nm + m^2) / d_{\rm R}$ with
$d_{\rm R} = \gcd(2n + m, 2m + n)$
	\cite{Izumida2016, Izumida2015}.
There is an arbitrariness for the choice of $p$ and $q$
($\bm{R}$ can be added by integer times $\bm{C}_{\rm h}/d$),
which does not change $a_z$ but changes $\theta_R$.
\bibitem{note3:winding_number}
Mathematically, the topological argument is valid only when
the sublattice symmetry holds, that is,
$\epsilon^{(0)}_{\mu, s} (k) = 0$ in the whole Brillouin zone.
However, for an insulator with
$|\epsilon^{(0)}_{\mu, s} (k)| < |f_{\mu, s} (k)|$,
the winding number is well-defined.
\end{thebibliography}
\end{document}